\documentstyle[preprint,prc,aps,psfig,,eqsecnum]{revtex}
\begin{document}
\draft

\title{Cross sections for the excitation of isovector charge-exchange
       resonances in $^{208}$Tl}

\author{G. Col\`o$^{\rm 1}$,
S. M. Lenzi$^{\rm 2}$,
E. E. Maqueda$^{\rm 3}$
and A. Vitturi$^{\rm 2}$}

\address{$^{\rm 1}$ Dipartimento di Fisica, Universit\`a degli Studi
and INFN, Sezione di Milano, \\Via Celoria 16, 20133, Milano, Italy}

\address{$^{\rm 2}$ Dipartimento di Fisica, Universit\`a degli Studi
and INFN, Sezione di Padova, \\Via F.Marzolo 8, 35131, Padova, Italy}

\address{$^{\rm 3}$ Departamento de F{\'{\i}}sica, CNEA, Avda. Gral.
Paz 1499, (1650) San Mart{\'\i}n, Argentina \\ and Consejo Nacional de
Investigaciones Cient{\'\i}ficas y T\'ecnicas}

\date{\today}

\maketitle

\begin{abstract}

The Glauber approximation for the treatment of heavy-ion scattering, has
already been shown to give reliable predictions for the reaction cross
section in the particular case of intermediate energy charge-exchange
processes. In the present work, we couple a Glauber-type model to
microscopic Random Phase Approximation calculations of the charge-exchange
excitations of $^{208}$Pb. The aim is to solve the longstanding question
whether the very elusive charge-exchange isovector monopole has been
really identified in the past experiments, or other multipoles were
prevalent in the observed spectra.  

\end{abstract}


\section{Introduction}
\label{sec:intro}

While some systematics is available for {\em isoscalar} giant resonances in
atomic nuclei, many {\em isovector charge-exchange} modes still remain
elusive. A number of intermediate energy (i.e., at bombarding energies 
E$_{lab}$ larger that 100 
MeV/A) reactions favour the excitation of spin-flip modes due to the 
dominance of the $\vec\sigma\cdot\vec\sigma$ part of the effective nucleon-nucleon 
interaction. For selective excitation of the non spin-flip modes, either pions 
or low-energy reactions or special projectile-ejectile combinations are needed.
Pioneering experiments of the first two types (cf., respectively, 
Refs.~\cite{Ere86} and~\cite{For87}) showed resonance-like structures, but the 
multipolarity assignments were difficult or impossible. This contrasts with the
abundance of theoretical predictions for the charge-exchange
giant resonances obtained in particular within the framework of 
self-consistent Random Phase Approximation (RPA), starting from the work of 
N. Auerbach and A. Klein~\cite{Aue83}. The lack of knowledge contrasts as 
well with the importance of charge-exchange isovector resonances for many
purposes. The isovector giant monopole resonance (IVGMR) is related to the
isospin impurity of the nuclear ground state~\cite{Aue_rep}, 
and to the isovector nuclear incompressibility~\cite{Won82} 
-- a nuclear matter parameter which is essentially unknown. 
More generally, the study of the excitations which involve the isospin 
degree of freedom, should give the possibility to test the nuclear models 
and eventually improve their predictive power for extreme conditions
(like nuclei with large isospin values, or nuclear matter close to isospin 
instability).

The most recent experiments aimed to the identification of the IVGMR and/or 
other multipole resonances, have employed the ($^{13}$C,$^{13}$N) reaction.
This reaction is believed to have good selectivity for the non spin-flip 
type of excitation in the target, because of the dominance of Fermi over 
Gamow-Teller transition in $^{13}$C. In the first experiment performed in
GANIL~\cite{CB} at 50 MeV/A, the candidate for the monopole excitation
in $^{208}$Tl (this is the $\Delta T_z$=+1 excitation of the $^{208}$Pb 
target) showed up at 21 MeV~\footnote{The excitation energies quoted in 
this work, are referred to the ground state of $^{208}$Pb, which means
that to get the actual excitation energy in $^{208}$Tl one has to 
subtract 4.2 MeV.}, 
in large disagreement with the result of 11 MeV found in the 
pion experiment of Ref.~\cite{Ere86}. One has to remember the 
severe problems of this pion experiment (low statistics, large background, 
full width of the monopole peak comparable to its energy), but also in the 
($^{13}$C,$^{13}$N) experiment there was the puzzling feature of the angular 
distribution which was inconsistent with $\Delta L$=0. In a second 
experiment~\cite{Lhe_two} at 60 MeV/A, the neutron decay of the monopole 
candidate at 21 MeV was measured, with the hope that an isotropic angular
distribution for the emitted neutrons could support the multipolarity
assignment. However, the results from this measurement are still ambigous,
as we discuss in more detail in Sec.~\ref{sec:results}.

Concerning the comparison with theory, the calculations of Refs.~\cite{Aue83,Col96} 
reveal a marked discrepancy of self-consistent RPA calculations with the 
results of the ($^{13}$C,$^{13}$N) experiment. 
According to Ref.~\cite{Col96}, the monopole peak is sensitive
to the effective force employed (a Skyrme parametrization), and it
is found either at 13 or 16 MeV, therefore much lower than
the energy region in which the experimental candidate for IVGMR is found.
On the other hand, in the same paper it was pointed out
that in that energy region
the RPA calculations predict a sizeable amount of octupole strength. 
A calculation of the relative 
values of $\Delta L$=0 and $\Delta L$=3 cross sections is therefore highly 
desirable. This further step of calculating the cross sections (instead of
simply limiting the analysis to the strength distributions) finds
its justification in the fact that the former do not obviously scale as the corresponding
strengths. 
Simple reaction frameworks that can however directly exploit
the structure information provided by the transition densities are
available in the literature, as for example
the one offered by the Glauber model. As shown in Ref.~\cite{MLVZ},
the extension of the Glauber model to the 
the description of heavy-ion charge-exchange reactions at intermediate 
energy, like in the case at hand,  can be 
rather successful in terms of comparison with experimental data. This 
motivates the present calculation of the cross sections for electric multipole 
charge-exchange excitations in the Glauber approach, at 60 MeV/A, 
having as ingredients microscopic transition 
densities of the RPA self-consistent calculation. 

We believe the present work may shed light on the experimental problem of
the IVGMR identification. We should mention that the efforts to improve
our rather poor knowledge of the various charge-exchange modes 
still continue in different laboratories~\cite{Tak_tbp}.  

The outline of the paper is the following. The formalism for the extension of the 
Glauber model to heavy-ion charge exchange reactions is briefly recalled in
Sec.~\ref{sec:cex}. The RPA calculations for the different
multipolarities are presented in Sec.~\ref{sec:target_densities}, where
they are discussed in particular in connection to
the coupling to the continuum and to the two particle-two hole states.
The predictions of the model for the charge-exchange cross sections are
presented and discussed in Sec.~\ref{sec:results}, with special emphasis 
on the energy range selectively populated by each multipolarity.
The main conclusions of the paper, as well as the implications for 
future experimental and theoretical work, are finally given.

\section{Charge exchange cross sections}
\label{sec:cex}

High energy scattering processes are dominated by nucleon--nucleon
collisions and can therefore be well reproduced by the Glauber
model~\cite{Gl}. In this approximation, nucleus--nucleus reactions are
fully microscopically described in terms of the nucleon-nucleon
scattering amplitudes and nuclear (transition) densities. The model was
developed in detail for charge-exchange reactions in a recent
paper~\cite{MLVZ} where the formalism is presented and is applied to
$p$-shell nuclei. For the sake of completeness we recall here some of
the main features and formulae relevant for the present work.

We assume that at relatively high energies these reactions are dominated by one
step processes~\cite{le2}.  Within this approximation, the scattering
amplitude of one--step charge-exchange reactions ($Aa \rightarrow
Bb$), is expressed as,
\begin{equation}
\label{ampli}
	f_{Aa \rightarrow Bb}(\Delta) = i k \sum_{LM} \int b~db~ 
\mu_{LM}^{Aa \rightarrow Bb}(b)~
			J_M(\Delta b)~e^{\lambda(b)+\chi(b)}~
e^{-iM\varphi_{_\Delta}},
\end{equation}
where $\Delta$ is the transferred momentum, $k$ is the
nucleus--nucleus relative momentum, $b$ is the impact parameter,
$\lambda(b)+\chi(b)$ represents the phase shifts including the nuclear
and Coulomb contributions and $J_M(\Delta b)$ is the Bessel function of
order $M$. The matrix element $\mu_{LM}^{Aa \rightarrow Bb}(b)$ takes into
account the details of the process and in Ref.~\cite{MLVZ} a general
expression for any type of charge-exchange reactions was given. Since
in the present case we will be only considering non-spin-flip transitions 
($\Delta S$=0), we cast the expression of $\mu_{LM}$ (Eq.~(2) of
Ref.~\cite{MLVZ}) in the form,
\begin{eqnarray}
\label{mu}
\mu_{LM}^{Aa \rightarrow Bb}(b) &=& 
 \frac{1}{ik_{NN}} 
	\langle~t_a~n_a~t_b~-n_b~\mid~1~n_a-n_b~\rangle
	\langle~t_A~n_A~t_B~-n_B~\mid~1~n_A-n_B~\rangle \nonumber \\
& & \sum_{J_p J_t} 
	\langle~j_a~m_a~j_b~-m_b~\mid~J_p~m_a-m_b~\rangle
	\langle~j_A~m_A~j_B~-m_B~\mid~J_t~m_A-m_B~\rangle \nonumber \\
& &\left[B_{J_p}~B_{J_t}\right]^L_M~~
 \int 
q~dq~   \hat{\rho}_{ab}^{J_p}(q)~\hat{\rho}_{AB}^{J_t}(q)~
f_{NN}^{10}(q)~J_M(qb),
\end{eqnarray}
where $k_{NN}$ is the nucleon--nucleon relative momentum. The above
expression includes the scattering amplitude $f_{NN}^{10}$, related to
the isovector central part of the effective nucleon--nucleon potential
and the Fourier transforms $\hat{\rho}(q)$,
of the projectile and target transition
densities $\rho_{ab}^{J_p}(r)$ and $\rho_{AB}^{J_t}(r)$, respectively. The
quantities $B$ arise from the projection of the spherical harmonics on
the plane perpendicular to the trajectory
($Y_{\ell}^m(\frac{\pi}{2},\varphi) = i^{\ell+m} B_{\ell m}
e^{im\varphi} \delta_{\ell+m,even}$). The squared brackets are used to
indicate the coupling of the associated angular momenta.

From the Eqs.~(\ref{ampli}) and (\ref{mu}) we can straightforwardly
obtain the differential cross section to each final state,
\begin{eqnarray}
\label{dsigma}
\lefteqn{
\frac{d\sigma}{d\Omega}(Aa \rightarrow Bb,\theta) =}& &  \nonumber \\
& &\frac{1}{\hat{J_a} \hat{J_A}}~\left( \frac{k}{k_{NN}} \right)^2 
\langle~t_a~n_a~t_b~-n_b~\mid~1~n_a-n_b~\rangle^2~
\langle~t_A~n_A~t_B~-n_B~\mid~1~n_A-n_B~\rangle^2   \nonumber \\
& &\sum_{J_p J_t} \sum_{L M} 
\left| [B_{J_p} B_{J_t}]^L_M               
\int b~db~e^{\lambda(b)+\chi(b)}~J_M(\Delta b)~
e^{-iM \varphi_{\Delta}}
\int q~dq~\hat{\rho}^{J_p}~\hat{\rho}^{J_t}~f_{nn}(q b)
J_M(qb) \right|^2, 
\end{eqnarray}
and, integrating over the scattering angle, the corresponding total cross section
\begin{equation}
\label{sigtot}
\sigma_{Aa \rightarrow Bb} = \int~b ~db ~
\left| e^{\lambda(b)+\chi(b)}~\mu_{LM}^{Aa \rightarrow Bb}(b) \right|^2.
\end{equation}

The nucleon-nucleon scattering amplitudes have been taken from the tabulation given in
Ref.\cite{LF} and interpolated for the energy of interest, i.e., 60 MeV/A.
The transition densities corresponding to projectile (ejectile)
have been obtained as in Ref.\cite{MLVZ} using the wave functions that are
solution of a shell model calculation in the $p$-shell~\cite{CK}. Due
to the non spin-flip character of the processes involved, only
transitions with $J_p = 0$ are allowed for the ($^{13}$C,$^{13}$N)
reaction 
regardless of the states populated in
$^{208}$Tl. The calculation
of the other crucial ingredients, i.e., the transition densities 
to each excited state in the target, is discussed in the next Section.

\section{MICROSCOPIC TRANSITION DENSITIES FOR TARGET EXCITATION}
\label{sec:target_densities}

As mentioned already, the transition densities corresponding to $^{208}$Pb
excitations are derived from charge-exchange RPA using Skyrme effective
forces. The method is well known, and the results for the IVGMR in $^{208}$Tl
have been published in Ref.~\cite{Col96}. We recall however the main features of
this type of calculations and we report some detail of the present one.

We choose the Skyrme parametrization SIII~\cite{Bei75}. In Ref.~\cite{Col96}, 
the sensitivity to the choice of the Skyrme parametrization is discussed
in the IVGMR case. We come back to this point in the next Section. In our
method, we first solve the Hartree-Fock (HF) equations in real space and we 
obtain the mean field. The unoccupied states, including those at positive
energies, are obtained by diagonalizing the mean field on a harmonic 
oscillator basis (in this case $\hbar\omega$ = 6.2 MeV), that is, the 
continuum is discretized. Using these single-particle states, a basis of
proton particle-neutron hole plus neutron particle-proton hole configurations
is built, and the RPA matrix equations are solved on this basis (which is
large enough to ensure that the appropriate sum rules~\cite{Aue83} are
satisfied). Our charge-exchange RPA is self-consistent, in the sense that the 
residual interaction between p-h configurations is derived properly from
the Skyrme force. Only the spin-orbit residual two-body force is dropped. 
The number of configurations used in the solution for the
different multipolarities (that is, $\Delta J$=$\Delta L$=0,1,2 and 3) is
provided in Table~\ref{tab:nstates}. It corresponds of course to the number of
RPA states: all these states have been included in the calculation of the
cross sections, that is, no selection based on their energy and/or strength 
has been applied. We present the results associated with the excited states
of $^{208}$Tl. In this case, the states have good isospin $T_0+1$ ($T_0$ is
the g.s. isospin of $^{208}$Pb) and it is possible to use them within the
framework of our Glauber model whose formulation includes the isospin quantum
number.  

Solving the RPA gives of course the energies $E_n$ of the excited states 
$\vert n\rangle$, as well as their wave functions in terms of the 
$X^{(n)}$ and $Y^{(n)}$ amplitudes. The corresponding
radial transition densities (mentioned above after Eq.~(\ref{mu})) 
are given by
\begin{equation}
\rho^{J_t}_n(r)=\sum_{ph}
    (X^{(n)}_{ph}-Y^{(n)}_{ph}) \langle p \vert\vert Y_L \vert\vert h \rangle
    R_p(r)R_h(r),
\end{equation}
where $ph$ labels the particle-hole components of the basis and $R(r)$
is the radial part of the single-particle wave function. These transition
densities are related to the strength of each state $S_n$ associated with 
the multipole operators $\hat O \equiv \sum_i r_i^L Y_L(\hat {r_i}) t_{\mu}^{(i)}$ by
\begin{equation}
S_n = \vert \int dr\ r^{L+2} \rho^{J_t}_n(r)\vert^2.
\end{equation}
The quantity $\rho^{J_t}_n(r)$ must be multiplied by a factor 
$\sqrt{{2\over 3}{2T_0+3\over 2J+1}}$ in order to be consistent with 
Eqs.~(\ref{mu}) and (\ref{dsigma}). 

\begin{figure}
\centerline{\psfig{figure=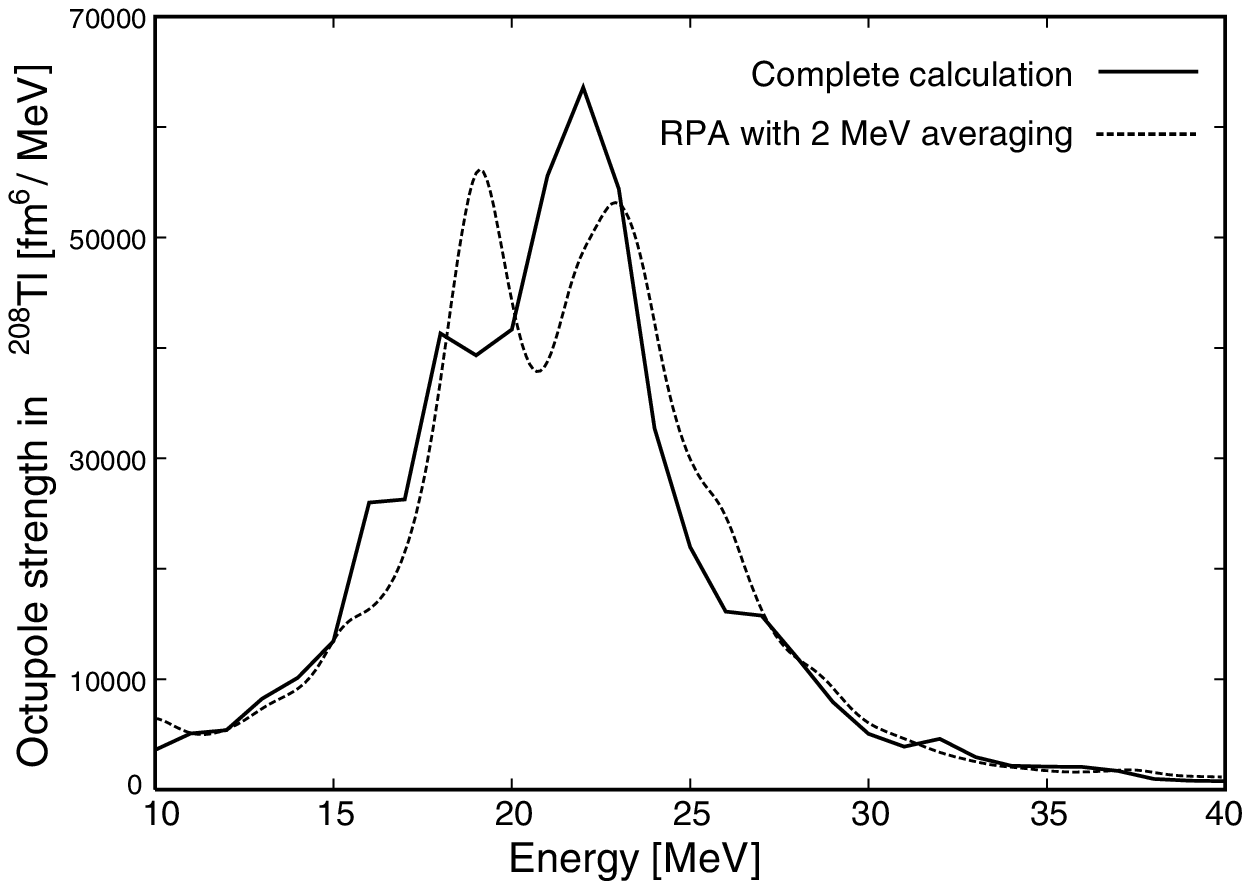,height=10.0cm}}
\vspace{5ex}
\caption{Distribution of charge-exchange octupole strength in $^{208}$Tl. A complete
calculation including escape and spreading (full line) is compared with the averaging
of the RPA results with Lorentzian functions having 2 MeV width (dashed line).}
\label{fig:octupole}
\end{figure}

In Ref.~\cite{Col96} more sophisticated calculations were performed. Using
the discrete RPA as a starting point, it is possible to take into account
the coupling with the continuum as well as with configurations of 
two particle-two hole type using the model first introduced in Ref.~\cite{Col94}.
Including the continuum coupling is equivalent to solve the well known
continuum-RPA. Both couplings have as a consequence a shift of the RPA peak(s), 
and also a redistribution of the strength whose main effect is to provide the
escape and spreading widths to the RPA peak(s). In the case of the IVGMR in 
$^{208}$Tl, as discussed in great detail in Sec. III of Ref.~\cite{Col96}, 
the discrete RPA peak is shifted downwards by about 2.8 MeV, from 19 MeV to
16.2 MeV. This effect is not included in the present calculation, therefore
the peaks corresponding to the 0$^+$ cross section in $^{208}$Tl should be
moved towards lower energies. This reinforces the final message of the
present paper, namely that the octupole cross section dominates in the region
around 21 MeV. In fact, we have checked that the octupole peak is much less 
shifted by the mentioned couplings. To substantiate this point, in 
Fig.~\ref{fig:octupole} we show the result of a full calculation of the octupole
strength, performed exactly as in~\cite{Col96} including escape and spreading, 
compared with the averaging of the discrete RPA results with 
Lorentzian distributions having 2 MeV width. The full calculation gives
essentially the same distribution as the discrete RPA + averaging.   

\section{Results for the cross sections and conclusions}
\label{sec:results}

The calculations described in previous Sec.~\ref{sec:target_densities} provide the
radial transition densities to all excited $0^+$, $1^-$, $2^+$, and
$3^-$ states in $^{208}$Tl. As indicated in Sect.~\ref{sec:cex}
these transition densities provide the basic ingredients for the calculation
of the charge-exchange cross sections. As an application,
we discuss here the results of our model for charge-exchange reaction
induced by $^{13}$C projectiles at 60 MeV/A on the $^{208}$Pb target.

\begin{figure}
\centerline{\psfig{figure=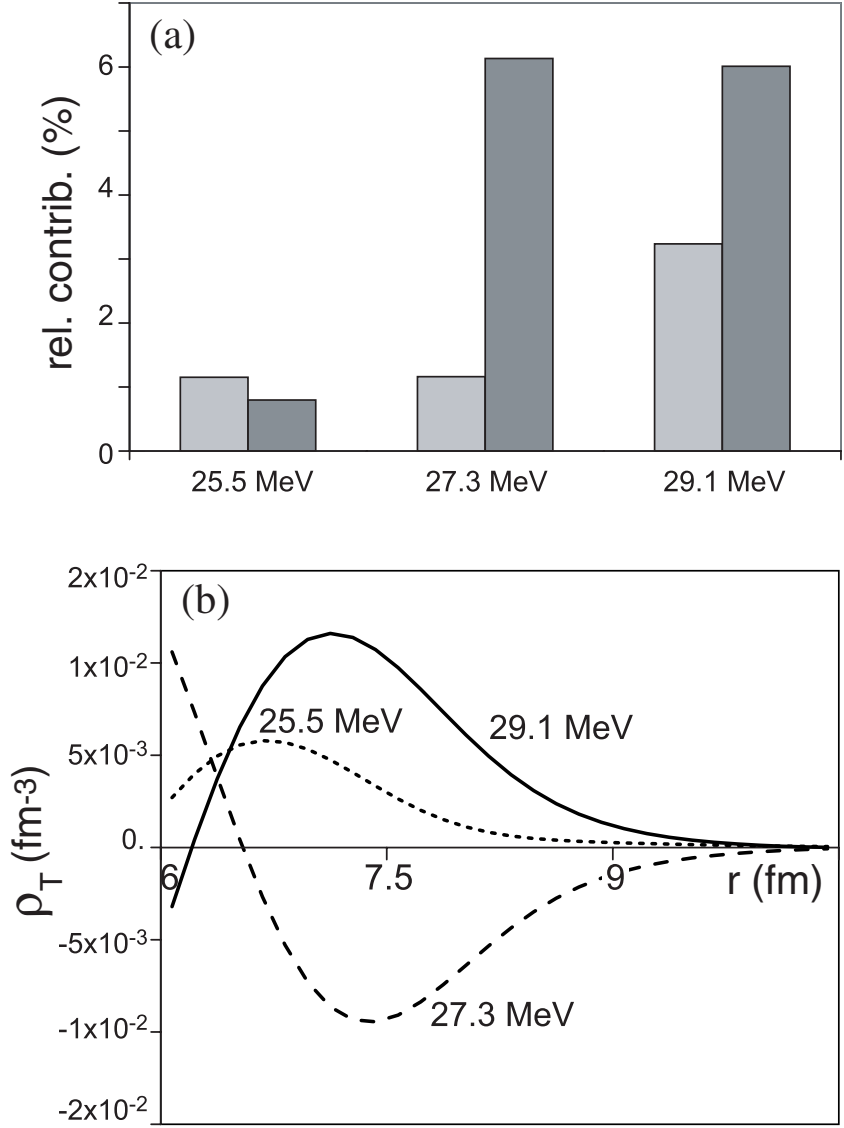,height=15.0cm}}
\vspace{5ex}
\caption{(a) comparison of the relative contributions, defined in terms of 
percentages of the total, to the dipole strength (light shadowed bars) 
and to the cross section (dark shadowed bars) in the
case of three different 1$^-$ states in $^{208}$Tl, whose energies are
reported in the figure. (b) Behavior
of the corresponding transition densities at the nuclear surface.}
\label{fig:colas}
\end{figure}

A first question to be posed, is whether the profile of the cross
sections as a function of the excitation energy follows that of the multipole
strengths. In Fig.~\ref{fig:colas}(a) we show some examples that
illustrate this point in the case of dipole states. We find that
in some cases states with similar dipole strength have rather different total
cross sections (as for example the states at
excitation energies $E_n$ = 25.5 and 27.3 MeV) and conversely, states with very
different strength have quite similar cross sections (as states at $E_n$ = 27.3 and 29.1
MeV). This fact can be understood by inspecting the radial behaviour of
the transition densities at the nuclear surface
[Fig.~\ref{fig:colas}(b)]. 
While the dipole transition matrix elements involve the transition densities
in the whole radial space, charge-exchange processes induced by heavy ions have a 
rather grazing character and the reaction amplitudes get their contributions mainly
from the surface region. We see therefore that the states
corresponding to the largest cross sections are those which have 
largest values for their transition densities at the surface, even if 
the associated strengths are not necessarily large.

\begin{figure}
\centerline{\psfig{figure=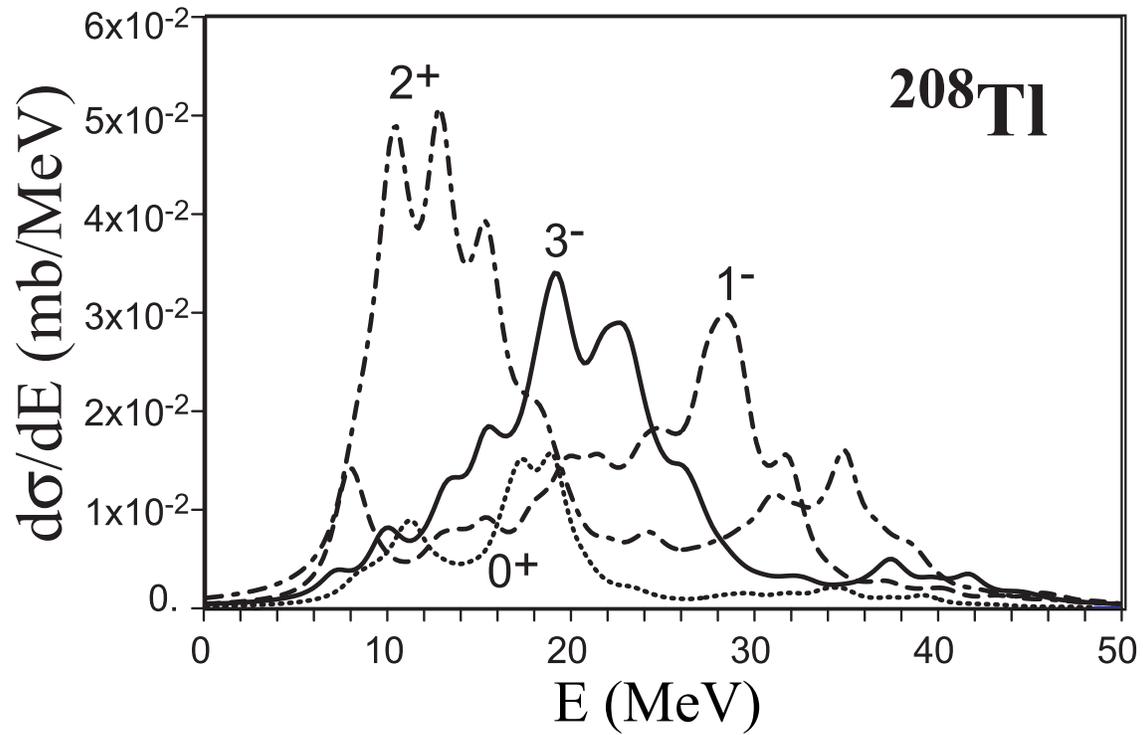,height=10.0cm}}
\vspace{5ex}
\caption{Total cross sections as a function of the energy in the
residual nucleus $^{208}$Tl for the different multipolarities.}
\label{fig:allmultipoles}
\end{figure}

On the basis of the above conclusion it is worth studying, as a
function of energy, the population of the different
states produced in a charge-exchange process, rather than to limit oneself to
the strength distributions. We have calculated the
angle integrated cross sections of Eq.~(\ref{sigtot}) for all the above
indicated multipoles.
In order to take into account the widths of the states we associate to each discrete 
RPA state a Lorentzian averaging
distribution as described in Sec.~\ref{sec:target_densities}, accounting for both
escape and spreading mechanisms. The widths of the Lorentzian distributions were 
chosen to be 2 MeV. The results are displayed in Fig.~\ref{fig:allmultipoles}. 
From the figure one can see 
that the main contributions correspond to
the quadrupole and octupole transitions in the energy range below 25 MeV. 

We finally discuss our predictions in more detail 
in connection with the experiment of Ref.~\cite{Lhe_two}. 
The aim of the experiment
was to infer the multipolarity of the observed structure from the
angular distribution of the neutron decay. From the excitation energy
spectra at small angles -- which in principle favour low multipole
transitions, in particular $\Delta L$=0 -- a large peak around 21 MeV is observed. 
The authors themselves point
out that the statistics for the neutron detection was too low to obtain a 
clear signature of
the multipolarity of this resonance from the neutron decay. Nevertheless, 
they attributed this peak to the
presence of the isovector giant monopole resonance. From the present
analysis we find that the octupole excitation should be dominant in the
21 MeV energy region, as shown in Fig.~\ref{fig:allmultipoles}
for the total cross
sections. This is in agreement with the remark made in~\cite{Col96} on the simple
basis of the strength distribution: nevertheless, we have seen that cross section
distributions and strength distributions can be different. 

\begin{figure}
\centerline{\psfig{figure=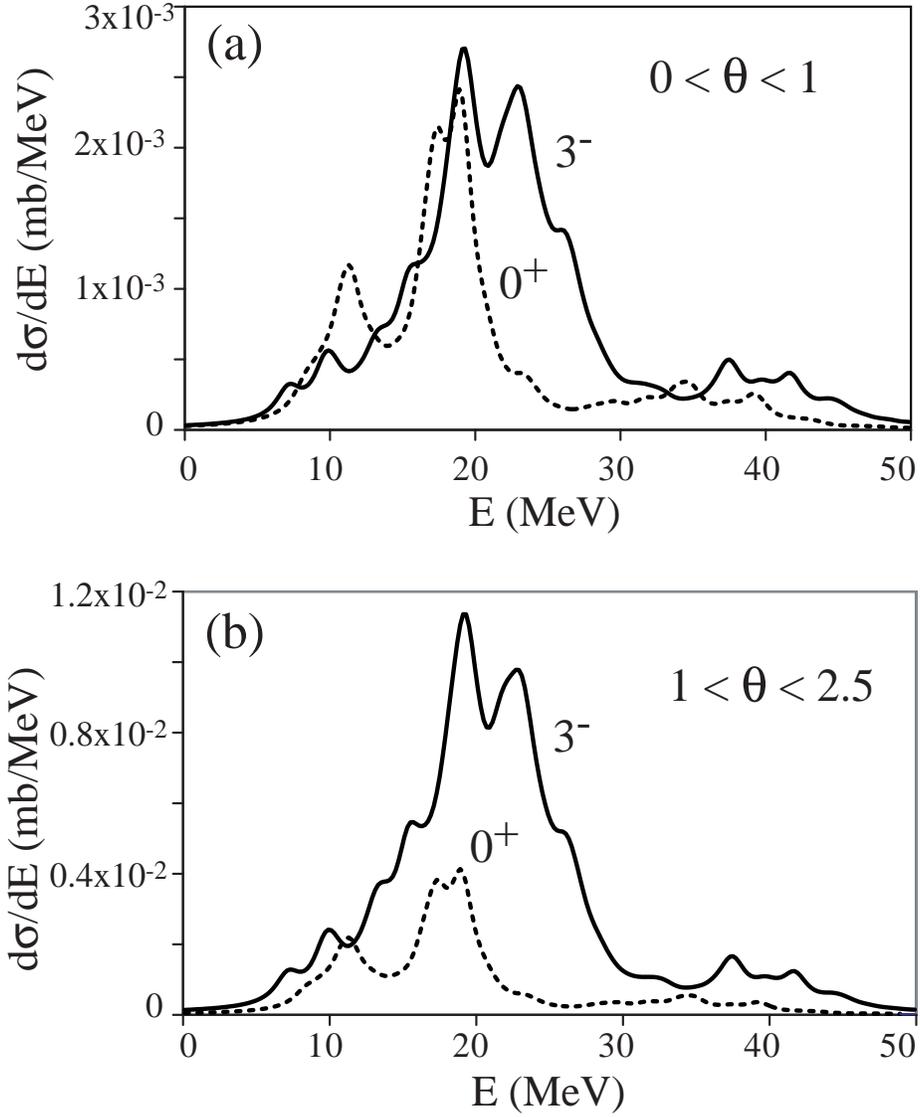,height=15.0cm}}
\vspace{5ex}
\caption{Cross sections as a function of the excitation energy for
monopole and octupole transitions in $^{208}$Tl integrated in the
angular range: (a) 0 to 1 degree, and (b) 1 to 2.5 degrees.}
\label{fig:diff_xsect}
\end{figure}

For small angles the relative importance of monopole
transitions increases. Therefore, a valid question is how the different
multipolarities would compare when only forward angles are considered.
In Ref.~\cite{Lhe_two} two different spectra are shown: one for the cross
section integrated between 0 and 1 degree and another integrated
between 1 and 2.5 degrees. Hence, we have calculated the excitation functions
for the monopole and octupole transitions by integrating the cross
sections of Eq.~(\ref{dsigma}) in the same angular range. The results
are shown in Fig.~\ref{fig:diff_xsect}. One can see that the relative
importance of the monopole with respect to the octupole excitations is
larger at small angles, but the octupole 
overcomes the monopole
transition in both angular ranges considered in~\cite{Lhe_two}.

In conclusion, the present paper shows the feasibility and usefulness of
fully microscopic calculations which couple structure and reaction
models. The Glauber model, which was already shown to be quite reliable for
the study of intermediate energy charge-exchange reactions, has been applied
to the case of 60 MeV/A ($^{13}$C,$^{13}$N) scattering on $^{208}$Pb. This
kind of experiment was peformed with the aim of extracting the properties of
the IVGMR. We have shown that the octupole cross section dominates over the
monopole in the energy range around 21 MeV, where the experimental peak was
identified, if one considers the angle-integrated total cross section. The ratio between the 
monopole and octupole cross sections grows if they are both integrated 
over small angular ranges; nevertheless, the monopole is not dominating in the angular
ranges considered in the experiment of Ref.~\cite{Lhe_two}. These remarks may
be of some importance when
the quest for the IVGMR, as well as for the other
still poorly known charge-exchange multipoles, 
continues.

\begin{table}
\caption{Number of particle-hole (ph) states included in the RPA calculation of the different
multipole excitations.}
\vspace{0.3cm}
\label{tab:nstates}
\begin{tabular}{ccc}
$J^{\pi}$ & neutron p-proton h & proton p-neutron-h\\\hline
$0^+$&64&88\\
$1^-$&172&240\\
$2^+$&244&348\\
$3^-$&288&420\\
\end{tabular}
\end{table}

\end{document}